\documentclass[pdflatex, sn-nature]{sn-jnl}% Style for submissions to Nature Portfolio journals

\usepackage{graphicx}%
\usepackage{multirow}%
\usepackage{amsmath,amssymb,amsfonts}%
\usepackage{amsthm}%
\usepackage{mathrsfs}%
\usepackage[title]{appendix}%
\usepackage{xcolor}%
\usepackage{textcomp}%
\usepackage{manyfoot}%
\usepackage{booktabs}%
\usepackage{algorithm}%
\usepackage{algorithmicx}%
\usepackage{algpseudocode}%
\usepackage{listings}%
\usepackage{xspace}
\usepackage{makecell}
\usepackage{bm}
\usepackage{enumitem}
\usepackage{cprotect}
\usepackage[normalem]{ulem}
\usepackage{float}
\theoremstyle{thmstyleone}%
\theoremstyle{thmstyletwo}%

\theoremstyle{thmstylethree}%

\renewcommand{\vec}[1]{{\bm{#1}}}
\newcommand{\oper}[1]{{\bm{\mathsf{#1}}}}

\newcommand{\software}[1]{{\textsc{#1}}}
\newcommand{\noisecov}[0]{\oper{C}^{-1}}

\newcommand{\dft}[0]{\oper{D}}

\newcommand{\lensop}[0]{\oper{L}}
\newcommand{\source}[0]{\vec{s}}
\newcommand{\regul}[0]{\oper{R}_\source}

\newcommand{\msol}[0]{\oper{A}}

\newcommand{\smp}[0]{\source_\mathrm{MP}}
\newcommand{\lams}[0]{\lambda_{\source}}
\newcommand{\data}[0]{\vec{d}}

\newcommand{\hyp}[0]{\mathrm{H}}
\newcommand{\etalens}[0]{\vec{\eta}_\hyp}

\newcommand{\logdet}[0]{\log \det}

\newcommand{\drr}[0]{\Delta\rr}

\newcommand{\lamg}[0]{\lambda_{\lpars}}
\newcommand{\rg}[0]{\oper{R}_{\lpars}}

\newcommand{\rr}[0]{\vec{r}}
\newcommand{\drrmp}[0]{\drr_\mathrm{MP}}

\newcommand{\lpars}[0]{\vec{\psi}}
\newcommand{\dlpars}[0]{\Delta\lpars}
\newcommand{\dsource}[0]{\Delta\source}

\newcommand{\xx}[0]{\vec{x}}

\newcommand{\lgi}[0]{\lensop_\mathrm{GI}}
\newcommand{\rgi}[0]{\oper{R}_\mathrm{GI}}

\newcommand{\Ds}[0]{\oper{G}_\source}
\newcommand{\Dpsi}[0]{\oper{G}_\lpars}
\newcommand{\defl}[0]{\vec{\alpha}}

\newcommand{\aodet}[0]{\mathcal{A}}
\newcommand{\vlbidet}[0]{\mathcal{V}}

\newcommand{\mmin}[0]{M_\mathrm{min}}
\newcommand{\mmax}[0]{M_\mathrm{max}}

\newcommand{\meighty}[0]{m_{80}}
\newcommand{\meightyv}[0]{m_{80,\vlbidet}}
\newcommand{\mfha}[0]{m_{400,\aodet}}

\newcommand{\mhm}[0]{M_\mathrm{hm}}

\newcommand{\msun}[0]{M_\odot}
\newcommand{\units}[1]{~\mathrm{#1}}
\newcommand{\dle}[0]{\Delta \log \mathcal{E}}

\newcommand{\rein}[0]{R_\mathrm{E}}
\newcommand{\sigmacrit}[0]{\Sigma_\mathrm{cr}}
\newcommand{\zs}[0]{z_\mathrm{s}}
\newcommand{\zl}[0]{z_\mathrm{l}}

\newcommand{\revisions}[1]{#1}
\newcommand{\secondrevisions}[1]{#1}

\newcommand{\pronto}[0]{\textsc{pronto}\xspace}

\newcommand{\bnte}[0]{JVAS B1938+666\xspace}

\newcommand{\znfs}[0]{SDSS J0946+1006\xspace}

\begin{document}

\title[]{\secondrevisions{A million-solar-mass object detected at cosmological distance using gravitational imaging}}


\author*[1]{\fnm{D.~M.} \sur{Powell}} \email{dmpowell@mpa-garching.mpg.de}

\author[2,3,4]{\fnm{J.~P.} \sur{McKean}} %\email{mckean@astro.rug.nl}

\author*[1]{\fnm{S.} \sur{Vegetti}}\email{svegetti@mpa-garching.mpg.de}

\author[5]{\fnm{C.} \sur{Spingola}} %\email{cristiana.spingola@inaf.it}

\author[1]{\fnm{S.~D.~M.} \sur{White}} %\email{swhite@mpa-garching.mpg.de}

\author[6]{\fnm{C.~D.} \sur{Fassnacht}} %\email{cdfassnacht@ucdavis.edu}

\affil[1]{\orgname{Max Planck Institute for Astrophysics}, \orgaddress{\street{Karl-Schwarzschild-Stra\ss{}e 1}, \postcode{85748} \city{Garching bei M\"unchen},  \country{Germany}}}

\affil[2]{\orgdiv{Kapteyn Astronomical Institute}, \orgname{University of Groningen}, \orgaddress{\street{PO Box 800}, \postcode{NL-9700 AV} \city{Groningen}, \country{The Netherlands}}}

\affil[3]{\orgname{South African Radio Astronomy Observatory}, \orgaddress{\street{PO Box 443}, \postcode{1740} \city{Krugersdorp}, \country{South Africa}}}

\affil[4]{\orgdiv{Department of Physics}, \orgname{University of Pretoria}, \orgaddress{\street{Lynnwood Road}, \postcode{0083} \city{Pretoria}, \country{South Africa}}}

\affil[5]{\orgdiv{Istituto di Radioastronomia}, \orgname{INAF}, \orgaddress{\street{via Gobetti 101}, \postcode{I-40129} \city{Bologna}, \country{Italy}}}

\affil[6]{\orgdiv{Department of Physics and Astronomy}, \orgname{University of California, Davis}, \orgaddress{\street{1 Shields Avenue}, \postcode{95616} \city{Davis}, \country{CA, USA}}}


\abstract{\textbf{Structure on sub-galactic scales provides important tests of galaxy formation models and the nature of dark matter. However, such objects are typically too faint to provide robust mass constraints.  Here we report the discovery of an extremely low-mass object detected via its gravitational perturbation to a thin lensed arc observed with milli-arcsecond-resolution very long baseline interferometry
(VLBI).  The object was identified using a non-parametric gravitational imaging technique, and confirmed using independent parametric modeling. It contains a mass of $\meighty=(1.13 \pm 0.04)\times 10^6\units{\msun}$ within a projected radius of 80 parsecs at an assumed redshift of 0.881. This detection is extremely robust and precise, with a statistical significance of $26\sigma$, a 3.3 percent fractional uncertainty on $\meighty$, and an astrometric uncertainty of 194 $\mu$as.  This is the lowest-mass object known to us, by two orders of magnitude, to be detected at a cosmological distance by its gravitational effect. This work demonstrates the observational feasibility of using gravitational imaging to probe the million-solar-mass regime far beyond our local Universe.
}}

% The cold dark matter (CDM) paradigm predicts the existence of dark matter haloes spanning many orders of magnitude in mass, from cluster scales ($>10^{13}\units{\msun}$) down to Earth-mass ($\sim10^{-3}\units{\msun}$) \cite{wang2020}.  While CDM has been remarkably successful at describing large-scale properties of the Universe, the evidence for CDM at mass scales below $\sim10^8\units{\msun}$ is less robust \citep{bullock2017}, largely due to the inherent difficulty of probing such low-mass structures from a cosmological distance. Here w

% \keywords{Strong gravitational lensing, VLBI, Galaxies, Dwarf galaxies, Globular clusters, Dark matter}


\maketitle

 On sub-galactic scales, observational evidence for the $\Lambda$CDM cosmological paradigm has been inconclusive due to the complex and highly non-linear interplay between baryons, radiation, and dark matter \cite{bullock2017}. A major challenge for conducting robust observational tests of $\Lambda$CDM in the low-mass regime ($\lesssim 10^8\units{\msun}$) is the inherent difficulty in characterizing the luminosities, sizes, and total masses of dim objects at large distances.  Low-mass luminous objects, such as ultra-compact dwarf galaxies and globular clusters, have so far only been directly observed within the local Universe out to $z\sim0.4$ \cite{phillipps2001,drinkwater2003,hasegan2005,mieske2008,lee2022}, while studies leveraging strong lensing by galaxy clusters have detected dwarf galaxies \cite{karman2017}, star clusters \cite{vanzella2022,mestric2022,adamo2024}, and even individual stars \cite{welch2022,diego2023} up to $z\sim10$. However, such observations are only sensitive to the emitted light, and any constraints on the total mass of such objects must be inferred indirectly.  Other low-mass objects are intrinsically dark, with dark matter haloes \cite{bertone2005,ackermann2015} and isolated intermediate-mass black holes (IMBH; \cite{bellovary2010}) expected to have no observable electromagnetic (EM) signatures.

% GCs out to $z=0.39$ \cite{lee2022}.

Strong gravitational lensing offers a powerful alternative pathway for studying low-mass objects with little to no EM luminosity. A spatially extended source in a galaxy-scale strong lens system acts as a backlight for the gravitational landscape of its lens galaxy, revealing low-mass perturbers through their gravitational effect alone. Furthermore, lens galaxies typically lie in the redshift range \revisions{$0.2 \lesssim z \lesssim 1.5$ \cite{robertson2020}}, meaning that low-mass, low-luminosity objects can be detected and studied across cosmic time.  To date, observations of galaxy-scale lenses with resolved arcs have been used to detect three low-mass perturbers: A $\sim 10^9\units{\msun}$ detection in \znfs \cite{vegetti2010,minor2021,ballard2024,despali2024, minor2024,enzi2024, tajalli2025}, a $\sim 10^8\units{\msun}$ detection in \bnte \cite{vegetti2012, sengul2022, despali2024, tajalli2025} (the lens studied in the present work), and a $\sim 10^9\units{\msun}$ detection in SDP.81 \cite{hezaveh2016, inoue2016}.  These were discovered in observations with the \textit{Hubble Space Telescope} (HST; 120 mas FWHM angular resolution), the W. M. Keck Observatory adaptive optics system (Keck AO; 70 mas), and the Atacama Large Millimeter Array (ALMA; 23 mas), respectively. Angular resolution is the limiting factor in detecting low-mass perturbers \cite{despali2022}, as the mass of a perturber sets a characteristic gravitational lensing length scale that must be resolved for a detection to be possible \cite{minor2017, despali2024}.  Expanding the mass range that we can robustly probe therefore necessitates that we use strong lens observations at the highest possible angular resolution.

\bnte \cite{lagattuta2012} is a strong lens system consisting of an early-type galaxy (ETG) at $\zl=0.881$ \cite{tonry2000} lensing both optical and radio emission from a source galaxy at $\zs=2.059$ \cite{riechers2011}; see Figure \ref{fig:deconvolved}.  For this work, we analyzed a radio observation of \bnte performed using a global VLBI array at 1.7 GHz, with a resulting angular resolution of $\sim 5\units{mas}$ \secondrevisions{(McKean et al., in press)}. %\cite{mckean2025}.  
In the radio, \bnte contains a bright quadruply-imaged feature, two images of which merge into an extremely thin arc, about $200\units{mas}$ in length, as well as a fainter doubly-imaged component.

We use a visibility-plane Bayesian forward-modeling technique \citep{vegetti2009, rybak2015,rizzo2018,powell2021,powell2022} to jointly infer both the pixellated source surface brightness distribution and the lens mass model.  We initially model the lens mass distribution in a fully parametric fashion to infer the large-scale mass distribution of the lens galaxy (the ``macro-model'').  We additionally include a truncated isothermal (pseudo-Jaffe; PJ) parametric mass component for the previously-detected $\sim 10^8\units{\msun}$ perturber, which we label $\aodet$ hereafter.  The VLBI observation of \bnte is sensitive to the presence of $\aodet$, with a Bayesian log-evidence of $\dle=+16$ over a completely smooth macro-model containing no low-mass perturbers.  

To search for additional low-mass haloes in \bnte, we next apply the gravitational imaging (GI) technique \citep{koopmans2005,vegetti2009, vegetti2010,vegetti2012}, a pixellated, non-parametric method for finding local overdensities in strong lens systems. In addition to the parametric macro-model and perturber $\aodet$, we recover a single strong, compact, positive correction to the surface mass density along the bright extended arc \secondrevisions{(Figure \ref{fig:deconvolved})}.  The presence of this feature is robust to the choice of GI pixel size and Bayesian prior. We hereafter label this new detection as $\vlbidet$.

To confirm detection $\vlbidet$ using a complementary technique, we repeat the analysis using a fully parametric model for $\vlbidet$ \secondrevisions{(see Table \ref{tab:logevs})}.  We find that a PJ profile with free truncation radius is preferred over a lens model containing only the macro-model and $\aodet$ by $\dle=+348$, corresponding to a detection significance of $26\sigma$. The inferred parameters for this model are $m_{\mathrm{tot},\vlbidet} = (2.8 \pm 0.3)\times 10^6\units{\msun}$ and a truncation radius $r_{t,\vlbidet}=149 \pm 18\units{pc}$. \revisions{Since} $m_{\mathrm{tot},\vlbidet}$ and $r_{t,\vlbidet}$ are highly correlated, for a more physically meaningful parameterization of the properties of $\vlbidet$ we define $\meightyv$ to be the cylindrical mass of $\vlbidet$ enclosed within a projected radius of 80 pc on the lens plane. This is the radius at which the enclosed mass exhibits the smallest statistical uncertainty (equivalently, where the enclosed mass and $r_{t,\vlbidet}$ de-correlate completely), indicating that the VLBI observation of \bnte is primarily sensitive to $\meightyv$. We find $\meightyv=(1.13 \pm 0.04)\times 10^6\units{\msun}$. This measurement is robust across various model choices for $\aodet$ and $\vlbidet$, and is also consistent with the gravitational imaging results; across six different GI runs, the projected mass within a radius of 80 parsecs ranges from $8.3\times10^5\units{\msun} \leq \meightyv \leq 1.8\times10^6\units{\msun}$ \secondrevisions{(Figure \ref{fig:mencl})}.  The inferred position of $\vlbidet$ is extremely precise, with an uncertainty of $194\units{\mu as}$ in the $y$-direction and $86\units{\mu as}$ in the $x$-direction, corresponding to 1.5 pc and 0.7 pc at the lens redshift, respectively.

 Figure \ref{fig:modelcmp} shows a visual explanation of the detection mechanism.
 Examining a comparison of reconstructed sky models with and without detection $\vlbidet$, it is clear that $\vlbidet$ corrects a $> 5 \sigma$ peak in the residuals map. Without perturber $\vlbidet$, the source model attempts (unsuccessfully) to absorb a defect in the lens model, resulting in a discontinuity in the reconstructed image of the bright radio lobe.  When $\vlbidet$ is introduced in the lens model, the caustic curve folds back over onto the source, simultaneously removing the surface brightness discontinuity and improving the model residuals. Hence, the drastic improvement in $\dle$ in the presence of perturber $\vlbidet$ is the result of both improved $\chi^2$ and a reduced regularization penalty due to a better-behaved source reconstruction.

If perturber $\vlbidet$ is a dark matter halo, its presence is consistent with \revisions{the expected number of detectable subhaloes in a $\Lambda$CDM cosmology}. In a $1.06\times10^{-2}\units{arcsec^2}$ sensitive region \revisions{around all lensed images} (roughly one beam width), the probability of detecting at least one CDM subhalo in the $[10^6,10^7]\units{\msun}$ mass bin is $p(n \geq 1)=0.65$. This probability falls to $p=0.36$ for 9.1 keV thermal relic warm dark matter (WDM) and $p=0.14$ for 4.6 keV WDM.
\revisions{While  detection $\vlbidet$ presents no strong tension with CDM or WDM models in terms of halo abundance, its density profile may be problematic.  In a companion paper \secondrevisions{(Vegetti et al., submitted)}, a range of different parametric density profiles beyond PJ are fit to this detection and an analysis of its implications for cold, warm, and self-interacting dark matter models is presented.  Interestingly, the best-fit NFW subhalo model from that analysis has a concentration that is many $\sigma$ higher than the expectation based on simulated CDM subhaloes \cite{moline2023}, an intriguing result similar to that found for object $\aodet$ in \bnte and for the $\sim 10^9\units{\msun}$ object in \znfs (see \cite{despali2024} and \cite{tajalli2025}). }

% By matching $\meightyv=1.13\units{\msun}$ and the logarithmic slope of the projected PJ profile at $80\units{pc}$, $\frac{1}{\meightyv} \frac{\mathrm{d}m_\vlbidet}{\mathrm{d}r}\vert_{r=80\units{pc}} = 1.31$ to a proxy NFW profile, we obtain a concentration of XXXXX. This is XXX$\sigma$ away from the mean concentration of an NFW subhalo of this mass. 

\revisions{When considering profiles for objects other than dark matter haloes, \secondrevisions{Vegetti et al. (submitted)} find that parametric profiles for intermediate-mass black holes \cite{bellovary2010} and globular clusters \cite{he2018} are disfavored at high statistical significance, though a relatively extended ultra-compact dwarf galaxy is not ruled out \cite{seth2014}. }  A more definitive statement on what type of object $\vlbidet$ is will require deep \revisions{optical/infrared} observations to detect any potential EM emission, a task made more challenging by the lensed optical emission seen in the Keck AO image of \bnte (Figure \ref{fig:deconvolved}).

% , and ultra-compact dwarf galaxies \cite{seth2014} are all 

Given the robustness of detection $\vlbidet$ in the gravitational imaging analysis, in tandem with its extremely high significance in independent parametric modeling, we claim $\vlbidet$ as the first $10^6\units{\msun}$ object directly detected at a cosmological distance using only its gravitational lensing effect.  The precision measurement of its mass, size, and position is unprecedented for an object in this mass range at this distance.  This result was made possible by milli-arcsecond-resolution global VLBI observation, analyzed using state-of-the-art lens modeling software. This work establishes VLBI strong lens observations as the only currently viable tool for discovering and precisely measuring properties of low-luminosity objects in the $10^6\units{\msun}$ mass regime \revisions{at cosmological distances}.

\newpage
\section{Methods} \label{sec:methods}

\subsection{Observation} \label{sec:observation}

The VLBI observation of \bnte used for this work was performed using a global VLBI array combining antennas from the Very Long Baseline Array (VLBA) and European VLBI Network (EVN) at 1.7 GHz for 14 hours at a data recording rate of 512 Mbits s$^{-1}$ (ID: GM068; PI: McKean). The observing strategy followed a standard phase referencing mode only for the VLBA part (phase calibrator J1933+654). Throughout the observations, scans on the fringe finder 3C454.3 were also included every $\sim4$ hours for the bandpass calibration. The correlation of the data from the 19 antennas was performed at the Joint Institute for VLBI–European Research Infrastructure Consortium (JIV–ERIC), and resulted in a dataset with 8 Intermediate Frequencies (IFs), each with 8 MHz bandwidth and divided into 32 channels for two circular polarizations (RR and LL).

The data were processed within the Astronomical Image Processing System ({\sc aips}, \cite{greisen2003}) package following a typical calibration procedure for phase-referenced observations. The resulting deconvolved image was obtained with a restoring beam of $7.4\units{mas}\times4.7\units{mas}$ at a position angle of 32.1 degrees east of north \secondrevisions{(McKean et al., in press)}. %\cite{mckean2025}. 
The noise, assumed to be Gaussian and uncorrelated, was measured directly in the Fourier domain by subtracting time-adjacent visibilities to remove the source signal, then computing the RMS within 30-minute time intervals. The triangle of highly sensitive baselines between Effelsberg, Jodrell Bank, and Westerbork were flagged to prevent them from dominating the model. The baselines Green Bank - Hancock, Green Bank - Owens Valley, Green Bank - Pie Town, and Los Alamos - Pie Town were also flagged for several 1 to 2 hour time intervals due to strong radio frequency interference (RFI).

The surface brightness distribution of the background radio source consists of two radio lobes (hot spots), the brighter of which is lensed into an extended gravitational arc, while the fainter one is doubly imaged (see Figure \ref{fig:deconvolved}). The physical properties of the source are derived and discussed by \secondrevisions{McKean et al. (in press)}. %\revisions{by McKean et al. \cite{mckean2025}}. 

\subsection{Bayesian inference} \label{sec:pronto}

We use the visibility-plane gravitational lens modeling software \pronto \citep{vegetti2009, rybak2015,rizzo2018,powell2021,powell2022} to jointly infer the pixellated source surface brightness vector $\source$ and lens parameters $\etalens$ from the observed visibilities $\data$, as well as to obtain the Bayesian log-evidence for each model parameterization $\hyp$.  At each likelihood evaluation (the first level of inference), we obtain the maximum a posteriori source $\smp$ for a given $\etalens$ and source regularization weight $\lams$ by solving
\begin{equation} \label{eqn:leastsquares}
   \msol \, \smp = (\dft\lensop)^T \noisecov \data\,,
\end{equation}
where
\begin{equation} \label{eq:msol}
    \msol \equiv \left[(\dft\lensop)^T\noisecov \dft\lensop + \lams \regul \right]\,.
\end{equation}
Here, $\lensop(\etalens)$ is the lens operator, which maps light from the source plane to the lens plane, $\lams \regul$ is the source prior covariance, which enforces the lens equation by penalizing strong surface brightness gradients, and $\noisecov$ is the noise covariance of the data.
 We solve Equation \eqref{eqn:leastsquares} using a preconditioned conjugate gradient solver, where the Fourier operator $\dft$ is implemented using a nonuniform fast Fourier transform (NUFFT). We refer the reader to \cite{powell2021} for further details on the method.

In the second level of inference, we sample the lens parameters $\etalens$ and source regularization weight $\lams$. The posterior is 
\begin{equation}\label{eq:bayev}
P(\etalens,\lams\mid\data) = \frac{ P(\data\mid
		\etalens,\lams) \, P(\etalens) \, P(\lams)}{ P(\data \mid \hyp)}\,.
\end{equation}
 The likelihood (which is the evidence from the source-inversion step) is
\begin{multline} \label{eq:baypos}
    	2\log{P(\data \mid \etalens, \lams)} = -\chi^2 - \lams \smp^T \regul \, \smp - \logdet \msol \\
    	+ \logdet (\lams \regul)  + \logdet (2\pi \noisecov)\,.
\end{multline}
This expression follows from the marginalization over all possible sources $\source$ when the noise and source prior are both Gaussian. As $\regul$ and $\noisecov$ are sparse, the terms containing them are straightforward to evaluate.  Computing $\logdet\msol$ is non-trivial; we approximate it using the preconditioner from the inference on $\smp$ as described by \cite{powell2021}. The $\chi^2$ term is
\begin{equation} \label{eq:chi2}
    \chi^2 = (\dft\lensop\smp-\data)^T\noisecov(\dft\lensop\smp-\data).
\end{equation}
We optimize its evaluation using the fast-$\chi^2$ technique from \cite{powell2022}. 

We compare different model parametrizations, $\hyp$, using the log-evidence $\log \mathcal{E}_\hyp \equiv \log P(\data \mid \hyp)$, which is obtained by marginalizing over all parameters. We use the nested sampling algorithm \software{MultiNest} \cite{feroz2009} to compute the log-evidence and to sample the posterior distributions of $\etalens$ and $\lams$.

\revisions{
Our method takes both noise and priors to be Gaussian. From a practical standpoint, these assumptions make our analysis computationally tractable by allowing for the marginalization over all possible configurations of the source and pixellated potential corrections (Section \ref{sec:gi}) using a single linear solution.  While it is the case that non-Gaussian noise statistics can in principle arise from calibration errors or correlated instrumental effects, we do not expect this to be a major issue given the high SNR of the observation. We have nevertheless tested our analysis on several independent sets of calibration solutions, including on individual spectral windows, finding our results to be robust across all cases. Most reassuringly, we find that the gap in the bright arc induced by object $\vlbidet$ cannot be suppressed by the self-calibration process.}

\revisions{Similarly, while there exist techniques for applying highly informative generative machine-learning models as non-Gaussian priors for ``realistic'' sources and potentials \cite[e.g.][]{adam2023}, these are susceptible to systematic biases learned from the training data (which are themselves taken from numerical simulations). False-positive subhalo detections are a particularly relevant concern for this work.  We instead opt for highly uninformative Gaussian priors,  which are agnostic to the shapes of the source surface brightness and the linearized potential corrections, thus maximizing the ability of the data to drive the inference.
}

\subsection{Parametric lens model components} \label{sec:macro}

We now introduce the surface mass density profiles used for parametric modeling in this work. All quantities labeled $\kappa$ are in units of the critical surface mass density for strong lensing, 
\begin{equation}
    \sigmacrit = \frac{c^2 \, D_s}{4\pi \, G \, D_{ls} \, D_l},
\end{equation}
where $D_s$, $D_l$, and $D_{ls}$ are the angular diameter distances from observer to source, from observer to lens, and from lens to source, respectively. Assuming a Planck 2015 \cite{planck2015} cosmology, $\sigmacrit=1.50 \times 10^{11}\units{\msun/arcsec^{2}}$ for \bnte ($\zs=2.059$, $\zl=0.881$). 

We model the smooth galaxy-scale mass distribution (the ``macro-model'') as an elliptical power-law mass distribution (e.g. \cite{keeton2001}), with a projected surface mass density of

\begin{equation} \label{eq:spemd}
    \kappa(\xi) = \frac{3-\gamma}{2} \left(\frac{\rein}{\xi} \right)^{\gamma-1},
\end{equation}
which we define in terms of an effective Einstein radius $\rein$ and the elliptical radius $\xi^2\equiv x^2 q + y^2/q$, where $q$ is the axis ratio. $\gamma$ is the 3D logarithmic slope, where $\gamma=2$ is isothermal. The additonal parameters $x_0$, $y_0$, and $\theta_0$ set the position and orientation of the profile. We use \software{FASTELL} \cite{barkana1999} to compute the corresponding deflection angles. 

To account for non-ellipticity in the lens galaxy (e.g. diskiness or boxiness), we include multipole perturbations of orders $m=3$ and $4$, parameterized as follows:
\begin{equation}
    \kappa_m(r,\theta) = \left(\frac{r}{1\units{arcsec}} \right)^{-(\gamma-1)}\left[a_m \sin(m\theta)+b_m\cos(m\theta)\right],
\end{equation}
which we write in circular polar coordinates here for readability.  The coefficients $a_m$ and $b_m$ encode the amplitude and orientation of each multipole term, and $\gamma$ is fixed to that of the underlying elliptical power-law model (Equation \ref{eq:spemd}). \revisions{We use Gaussian priors for $a_m$ and $b_m$ with $\mu=0$ and $\sigma=0.01$, a choice motivated by the results of numerical simulations of massive elliptical galaxies \revisions{(e.g. \cite{kochanek2004})}, which is also consistent with previous lens modeling results \cite{powell2022,stacey2024,tajalli2025}.}  We also include an external shear term in the macro-model, with strength $\Gamma$ and direction $\theta_\Gamma$.

%\todo{discuss priors on multipole coefficients?}
% We impose a Gaussian prior of width $\sigma=0.01$ on $a_m$ and $b_m$. Our choice of prior is motivated by \cite{kochanek2004}, who note a typical amplitude of $\kappa_0 \sqrt{a_4^2+b_4^2} \sim 0.005$ from numerical simulations. The conversion to our units is not exact, as they assume an isothermal ($\gamma = 2$) density slope, but it is sufficient for our purposes.

For parametric modeling of the low-mass perturbers $\aodet$ and $\vlbidet$ \revisions{(see Table \ref{tab:logevs})}, we use the following spherically-symmetric mass profiles.  Models \verb|PJ_free| and \verb|PJ_tidal| are truncated isothermal profiles (Pseudo-Jaffe; e.g. \cite{jaffe1983,keeton2001}) with a projected surface mass density of
\begin{equation} \label{eq:pj}
  \kappa(r) = \frac{ m_\mathrm{tot} }{2 \pi \, r_t^2  \sigmacrit } \,\left( \frac{1}{x} -
    \frac{1}{\sqrt{x^2+1}} \right),
\end{equation}
where $r_t$ is the truncation radius, $x=r/r_t$, and $m_\mathrm{tot}$ is the total mass.  We leave $r_t$ as a free parameter for \verb|PJ_free|, while for \verb|PJ_tidal| we set $r_t$ to be the tidal radius 
\begin{equation} \label{eq:tidal}
    r_t = \frac{2 R}{3} \left ( \frac{m_\mathrm{tot}}{2 M_\mathrm{lens}(<R)} \right)^\frac{1}{3}.
\end{equation}
Here, $R$ is the 3D radius to the \revisions{centre} of the lens and $M_\mathrm{lens}(<R)$ is the mass of the main lens enclosed within $R$; see \cite{despali2024}.  Because we have no knowledge of the 3D position of a low-mass perturber within the lens, we take $R$ to be the projected distance to the perturber in the plane of the lens. 

In the case of \verb|NFW_sub|, we model perturber $\aodet$ as a spherical Navarro-Frenk-White (NFW) profile \cite{navarro1996, keeton2001}, with a surface mass density of
\begin{equation}
\kappa(x) = 2\kappa_s\frac{1-F(x)}{x^2-1}\,, 
\label{eq:nfw}
\end{equation}
where $x=r/r_s$, $\kappa_s = \rho_s r_s / \sigmacrit$, and 
\begin{equation}
F(x) = \begin{cases}
    \frac{1}{ \sqrt{x^2-1}}\,\tan^{-1} \sqrt{ x^2-1} &  x>1 \cr
    \frac{1}{ \sqrt{1-x^2}}\,\tanh^{-1}\sqrt{1-x^2} &  x<1 \cr
    1                                  & x=1 \cr
  \end{cases}.
\end{equation}

We have found that using a mass-concentration-redshift relation for cold dark matter haloes (e.g. \cite{duffy2008}) leads to virial concentrations that are much too low to produce the required lensing effect for a halo of a given virial mass (see also \cite{despali2024,tajalli2025}). \revisions{We therefore leave the concentration of} \verb|NFW_sub| \revisions{as a free parameter in this work.  The redshift is fixed to that of the main lens.}

\subsection{Gravitational imaging}
\label{sec:gi}

Gravitational imaging (GI; \cite{koopmans2005,vegetti2009, vegetti2010,vegetti2012,vernardos2022}) is a technique for recovering pixellated (non-parametric) corrections to the lensing convergence. GI is well-suited for finding local overdensities that have not been accounted for in the parametric mass model, $\etalens$, which remains fixed during the GI procedure.  Our GI formulation augments the lens operator (e.g. Equation \eqref{eqn:leastsquares}) with an extra block column representing linearized degrees of freedom in the lensing potential:
\begin{equation}
    \lensop_\mathrm{GI} \equiv \left[ \lensop \, |  \, -\lensop \Ds \Dpsi \right],
\end{equation}
where
\begin{equation} \label{eq:Ds}
    \Ds \equiv \frac{\partial \source}{\partial \xx} 
\end{equation}
is the matrix of source gradients, and
\begin{equation} \label{eq:Dpsi}
    \Dpsi \equiv \frac{\partial \defl}{\partial \lpars},
\end{equation}
is the matrix of derivatives of the deflection angles $\defl$ with respect to the pixellated lensing potential $\lpars$.  $\lgi$ acts on the concatenated vector of the pixellated source and linearized potential corrections:
\begin{equation}
     \rr \equiv \begin{pmatrix}\source\\ \lpars \end{pmatrix}, \;\; \drr \equiv \begin{pmatrix}\dsource\\ \dlpars \end{pmatrix}
\end{equation}
Similarly, we define the block regularization matrix
\begin{equation} \label{eq:rgi}
    \rgi \equiv \begin{pmatrix} \lams \regul & 0 \\ 0 & \lamg \rg  \end{pmatrix}.
\end{equation}
For this work, we apply three different forms for the potential regularization operator $\rg$, \revisions{checking them against one another to ensure the robustness of our results. These penalize total mass, surface density gradients, or surface density curvature. In each case, $\rg$ is assembled using standard finite-difference operators for the gradient and Laplacian on a Cartesian grid.}

It can be shown that the proper linearization over $\lpars$ yields a linear system analogous to Equation \ref{eqn:leastsquares}:
\begin{equation}  \label{eq:gisolve}
\left[(\dft \lgi)^T \noisecov (\dft \lgi ) + \rgi \right] \, \drrmp = - (\dft \lgi)^T \noisecov (\dft \lensop \source - \data)- \rgi \, \rr .
\end{equation}
Formally, Equation \eqref{eq:gisolve} represents a single Newton iteration in the space of $\source$ and $\lpars$. After iterating until convergence, it can be shown that the likelihood with respect to the regularization weights, $\log{P(\data \mid \lams, \lamg)}$ is obtained by substituting $\source\rightarrow\rr$, $\lams \regul \rightarrow \rgi$, and $\lensop\rightarrow \rgi$ in Equation \eqref{eq:baypos}.   A single likelihood evaluation in our GI procedure thus consists of:
\begin{enumerate}
    \item Initialize $\rr = \vec{0}$.
    \item \label{item:solve} Solve for $\drrmp$.
    \item \label{item:update} Update $\rr = \rr + \drrmp$.
    \item Repeat steps \ref{item:solve} and \ref{item:update} until convergence.  We find that 10 iterations are generally sufficient for convergence; in practice we use 20.
    \item Evaluate $\log{P(\data \mid \lams, \lamg)}$ using Equation \eqref{eq:baypos}.
\end{enumerate}
Note that our GI implementation differs from that of \cite{vernardos2022}, who use a single iteration per likelihood evaluation.  Our approach of iterating to convergence allows the GI algorithm to more easily recover compact features in the lensing potential. 

\revisions{We use a standard simplex optimization algorithm to maximize the likelihood with respect to the regularization weights $\lams$ and $\lamg$, where each likelihood evaluation requires 20 sub-iterations as described above.  We used starting values for $\lams$ and $\lamg$ of $10^{9}$ and $10^{13}$, respectively, with initial logarithmic step sizes of 0.5 dex. After optimizing for $\lams$ and $\lamg$, we convert $\lpars$ into corrections to the lensing convergence (surface mass density) via $\kappa_\mathrm{GI} = -\frac{1}{2}\nabla^2 \lpars$. }

\revisions{Our use of Gaussian priors for the linearized potential corrections was motivated by the need for an uninformative and computationally fast regularization term.  A natural consequence of this choice is that $\lpars$, and hence $\kappa_\mathrm{GI}$, prefer to resemble Gaussian random fields in regions where the data are uninformative.  To aid in our interpretation of the GI results, we apply a $3\sigma_\mathrm{GI}$ threshold to the convergence maps, treating any convergence over the threshold as a real feature.  We compute $\sigma_\mathrm{GI}$ from the RMS of the convergence corrections within the mask used for modeling the lensed images.  $\sigma_\mathrm{GI}$ must be empirically estimated in this way, as the operator-based iterative solver framework (Section \ref{sec:pronto}) precludes direct manipulation of the posterior covariance matrix. }

\subsection{Expected number counts of detectable subhaloes} \label{sec:counts}

We express the differential mass function for WDM subhaloes, in terms of the ``half-mode mass'' $\mhm$, as
\begin{equation}
\frac{dn}{dm} = m^{\alpha}\left[1+\left(\alpha_2\frac{\mhm}{m}\right)^\beta\right]^\gamma    
\end{equation}
with $\alpha = -1.9$, $\alpha_2 = 1.1 $, $\beta = 1.0$ and $\gamma =  -0.5$ \cite{springel2008,oriordan2022}. The mass $m$ is the total mass of a pseudo-Jaffe subhalo. The expected number of subhaloes in the mass range $\mmin=10^6\units{\msun}$ to $\mmax=10^7\units{\msun}$ is then
\begin{equation}
\mu_{\rm sub} = A_\mathrm{sens}\,f_\mathrm{sub}\,\frac{M_{\rm_{lens}}(<2\rein)}{4 \pi \rein^2}\frac{\int_{\ln{\mmin}}^{\ln{\mmax}}{m \frac{dn}{dm} \, \mathrm{d}\ln m}}{\int_{\ln{\mmin}}^{\ln{\mmax}}{m^{\alpha+2} \, \mathrm{d}\ln m}},
\label{eq:mu_sub}
\end{equation}
where $\rein$ is the Einstein radius of the lens, and $M_{\rm_{lens}}(<2\rein)$ is the total projected mass of the lens within twice the Einstein radius. $f_\mathrm{sub}$ is the fraction of dark matter (normalized with respect to the CDM mass function) contained in subhaloes; we use $f_\mathrm{sub}=0.012$ \citep{hsueh2020}. $A_\mathrm{sens}$ is the area around the lensed arcs that we expect to be sensitive to the presence of subhaloes, which we obtain by thresholding the deconvolved image at $5\sigma$ above the noise for a total area of $A_\mathrm{sens}=1.06\times10^{-2}\units{arcsec^2}$. This area roughly corresponds to one primary beam width ($\sim 5\units{mas}$) along each bright arc. Defining $A_\mathrm{sens}$ in such a tight region around the brightest parts of the lensed arcs is an intentionally conservative choice made to ensure that all subhaloes $\gtrsim 10^6 \units{\msun}$ within this region have been detected. \secondrevisions{Our use of a constant sensitivity within $A_\mathrm{sens}$ is a rough but conservative approximation to more sophisticated sensitivity mapping techniques (e.g., \cite{despali2022, oriordan2022}).}

\subsection{Parametric modeling: macro-model and $\aodet$ only}
\label{sec:parametric:a}

The results of the fully-parametric modeling are summarized in Table \ref{tab:logevs}.  We first consider those fully-parametric models consisting of the macro-model and the previously-detected perturber $\aodet$ (see Section \ref{sec:macro}).  \revisions{ Because $\aodet$ lies $\sim50\units{mas}$ away from the nearest lensed radio emission, its redshift cannot be robustly constrained by our VLBI observation alone. For this work, we therefore assume that $\aodet$ lies at the redshift of the main lens.}

\revisions{The best parametric models for $\aodet$, in terms of Bayesian log-evidence $\dle$, are truncated isothermal subhalo models (pseduo-Jaffe), with either free or fixed (tidal) truncation radii being equally preferred} ($\aodet=$ \verb|PJ_free| or $\aodet=$ \verb|PJ_tidal|; $\dle=16$). \revisions{We additionally tested an NFW profile with free virial concentration} ($\aodet=$ \verb|NFW_sub|; $\dle=13$), \revisions{as well as a completely smooth macro-model} ($\aodet=$ \verb|None|; $\dle\equiv0$). For simplicity, we choose the best model with fewer free parameters, $\aodet=$ \verb|PJ_tidal|, as our fiducial model for the gravitational imaging (GI) procedure.  We find that the choice of profile $\aodet$ does not impact the inferred properties of $\vlbidet$.

\subsection{Gravitational imaging detection of $\vlbidet$}

The results of the gravitational imaging (GI) modeling, applied to the parametric model with $\aodet=$ \verb|PJ_tidal|, are shown in \secondrevisions{Figure \ref{fig:deconvolved} and Supplementary Figure 1}. For three different regularization types (penalizing the convergence, gradient of convergence, or curvature of convergence) and two different grid resolutions ($N_\mathrm{GI}=512$ and $N_\mathrm{GI}=1024$, corresponding to pixel sizes of $\Delta x_\mathrm{GI}=3.5\units{mas}$ and $\Delta x_\mathrm{GI}=1.8\units{mas}$, respectively), we consistently recover a compact density correction well above the $3\sigma_\mathrm{GI}$ threshold used to identify \textit{bona fide} features in the surface mass density that were unaccounted for during the initial parametric modeling. We label this feature $\vlbidet$, as it was discovered using a VLBI observation of \bnte.  

Defining $\meightyv$ to be the cylindrical mass contained within a projected radius of 80 pc on the lens plane (see Section \ref{sec:parametric:all}), we find $8.3\times10^5\units{\msun} \leq \meightyv \leq 1.8\times10^6\units{\msun}$ for \revisions{the} six GI runs.  The scatter in GI masses is due to the large number of degrees of freedom in the pixellated convergence map, combined with the GI technique's lack of prior knowledge on the approximate sphericity of gravitationally-bound astrophysical objects. GI is an important technique for identifying and estimating the masses of positive density corrections to a parametric lens model; however, it is necessary to verify detection $\vlbidet$ with independent parametric modeling.

\subsection{Parametric modeling: macro-model, $\aodet$, and $\vlbidet$}
\label{sec:parametric:all}

We repeat the fully parametric modeling procedure, this time also including a parametric profile for perturber $\vlbidet$. The results are summarized in Table \ref{tab:logevs} and \secondrevisions{Supplementary Figure 3}.  We assume for this work that perturber $\vlbidet$ lies at the lens redshift of $z=0.881$, using a truncated isothermal profile (pseudo-Jaffe; see Section \ref{sec:macro}).  We find that, regardless of the choice of profile for detection $\aodet$, the inclusion of a parametric model for $\vlbidet$ increases the log-evidence by at least \revisions{$\dle>332$ ($>25\sigma$)} in all cases.  Models $\aodet=$ \verb|PJ_free|, $\vlbidet=$ \verb|PJ_free| and $\aodet=$ \verb|PJ_tidal|, $\vlbidet=$ \verb|PJ_free| are equally preferred. As before, we choose $\aodet=$ \verb|PJ_tidal|, $\vlbidet=$ \verb|PJ_free| with fewer free parameters for a more detailed discussion, and treat it as our ``best model'', with a significance of $\dle=364$ over the smooth macro-model and $\dle=348$ ($26\sigma$) over $\aodet=$ \verb|PJ_tidal|, $\vlbidet=$ \verb|None|.  For the rest of the discussion, it is implied that $\aodet=$ \verb|PJ_tidal|.

% \sout{The most-preferred model is $\aodet=$ \verb|NFW_LOS|, $\vlbidet=$ \verb|PJ_free| with $\dle=365$ over the smooth macro-model. However, for the reasons outlined in Section \ref{sec:parametric:a}, we omit field-halo interpretations of $\aodet$ from further discussion.}

In the $\vlbidet=$ \verb|PJ_free| parametrization, $\vlbidet$ has a total mass of $m_\vlbidet = (2.82 \pm 0.26)\times 10^6\units{\msun}$ and a truncation radius $r_{t,\vlbidet}=149 \pm 18\units{pc}$. To express the inferred mass in a way that is independent from $r_{t,\vlbidet}$, we find 80 pc to be the radius at which the enclosed (projected) mass $\meightyv$ is uncorrelated with $r_{t,\vlbidet}$. For $\vlbidet=$ \verb|PJ_free|, $\meightyv = (1.13 \pm 0.04) \times10^6\units{\msun}$, which is consistent with the gravitational imaging results (see Figure \ref{fig:mencl}). For $\vlbidet=$ \verb|PJ_tidal|, $\meightyv=(1.07 \pm 0.04)$ agrees with $\vlbidet=$ \verb|PJ_free|, indicating that the enclosed mass at 80 pc is indeed particularly well-constrained by the data.  The position of $\vlbidet$ is constrained to  $194\units{\mu as}$ precision in the $y$-direction and $86\units{\mu as}$ in the $x$-direction.  For all models, $\meightyv$, $x_\vlbidet$, and $y_\vlbidet$ are consistent within the $1\sigma$ uncertainties regardless of the choice of profile $\aodet$.  

The tidal truncation radius of $\vlbidet=$ \verb|PJ_tidal| (Equation \ref{eq:tidal}) is $r_{t,\vlbidet}=53 \pm 1\units{pc}$, a factor of 3 smaller than that inferred using \verb|PJ_free|. Relative to $\vlbidet=$ \verb|PJ_free|, $\vlbidet=$ \verb|PJ_tidal| is disfavored by $\dle=-16$, suggesting that this smaller value of $r_{t,\vlbidet}$ is too compact to produce the required lensing effect. This is likely a result of the use of the projected 2D separation as a proxy for the 3D distance between the lens galaxy and the perturber; any offset of $\vlbidet$ along the line of sight would increase the 3D distance and hence $r_t$, alleviating this issue. 

Unlike the Keck AO observation of \bnte, in which perturber $\aodet$ lies on top of the infrared arc, the projected lens-plane distance between $\aodet$ and the nearest lensed radio emission is 400 pc; we therefore expect only the cylindrical mass out to a projected radius of 400 pc, which we define as $\mfha$, to be well-constrained by the VLBI data. For $\aodet=$ \verb|PJ_tidal|, we find $\mfha=(5.0 \pm 0.8) \times 10^7\units{\msun}$ and $r_{t,\aodet}=243 \pm 20\units{pc}$. For $\aodet=$ \verb|PJ_free|, $\mfha=(5.2 \pm 0.7) \times 10^7\units{\msun}$ and $r_{t,\aodet}=15 \pm 20\units{pc}$. Hence, $\mfha$ is well-constrained and consistent between the two, while $r_{t,\aodet}$ is unconstrained and simply reflects the tidal radius relation and the log-uniform prior (respectively) used for these profiles. For comparison, Despali et al. (\cite{despali2024} and subsequent private communication) find a mass of $\mfha=(7.7 \pm 0.1)\times 10^7\units{\msun}$ for their best \revisions{subhalo} model of the Keck AO observation.

\newpage

\backmatter

\section*{Declarations}

\bmhead{Data availability} 

The VLBI dataset is publicly available on the EVN archive \verb|https://archive.jive.nl/scripts/portal.php| (Experiment GM068).  The Keck AO observation used in Figure \ref{fig:deconvolved} is publicly available on the Keck Observatory Archive \verb|https://koa.ipac.caltech.edu/| (Program ID U085N2L).

\bmhead{Code availability} 

\secondrevisions{The modeling code \pronto is not publicly available. The reader interested in using this code can contact \texttt{svegetti@mpa-garching.mpg.de}.}

\bmhead{Acknowledgments}

We thank Giulia Despali and Maryam Tajalli for helpful discussions on the Keck AO modeling results for \bnte. This research was carried out on the High-Performance Computing resources of the Raven cluster at the Max Planck Computing and Data Facility (MPCDF) in Garching, operated by the Max Planck Society (MPG). \revisions{The European VLBI Network is a joint facility of independent European, African, Asian, and North American radio astronomy institutes. The scientific results from the data presented in this publication are derived from the following EVN project code(s): GM068. The National Radio Astronomy Observatory is a facility of the National Science Foundation operated under cooperative agreement by Associated Universities, Inc.}

D.~M.~P. and S.~V. have received funding from the European Research Council (ERC) under the European Union's Horizon 2020 research and innovation programme (grant agreement No 758853). S.~V. thanks the Max Planck Society for support through a Max Planck Lise Meitner Group. C.~S. acknowledges financial support from INAF under the project “Collaborative research on VLBI as an ultimate test to $\Lambda$CDM model” as part of the Ricerca Fondamentale 2022. This work was supported in part by the Italian Ministry of Foreign Affairs and International Cooperation, grant number  PGRZA23GR03. This work is based on the research supported in part by the National Research Foundation of South Africa (Grant Number: 128943).

\bmhead{Author contributions statement}

D~M.~P. is the main developer of \textsc{pronto}, and has carried out the modeling and analysis presented here.  J.~P.~M. observed and reduced the VLBI data, and was actively involved in the writing process.  S.~V. contributed to the development of \textsc{pronto} and to the interpretation of the results. C.~S. calibrated the VLBI data and wrote the Observations section.  S.~D.~M.~W. contributed extensively to the interpretation of the results. C.~D.~F. observed the Keck AO data and contributed to the interpretation of the results.

\bmhead{Competing interests statement} 

We have no conflicts of interest to declare.

\newpage
\section*{Tables}

\begin{table}[h!]
\cprotect\caption{Properties of low-mass perturber $\vlbidet$ inferred from the fully parametric analysis, and their $1\sigma$ uncertainties.  $\vlbidet$ is fixed to the lens redshift $\zl=0.881$ in this work. The tidal radius $r_t$ of $\vlbidet=$ \verb|PJ_tidal| is derived from the lens and perturber masses and positions, and is not a free parameter.}\label{tab:logevs}
\begin{tabular}{llllll}
\toprule
Profile $\aodet$ & Profile $\vlbidet$ & $\dle$ &$m_{80,\vlbidet}$ ($10^6\units{\msun}$)  &  $m_{\mathrm{tot},\vlbidet}$ ($10^6\units{\msun}$) & $r_{t,\vlbidet}$ (pc)  \\ 
\midrule
\verb|PJ_tidal| & \verb|PJ_free| & 364 &  1.13 $\pm$ 0.04 & 2.82 $\pm$ 0.26 & 149 $\pm$ 18 \\
\verb|PJ_free| & \verb|PJ_free| & 364 &  1.14 $\pm$ 0.03 & 2.78 $\pm$ 0.23 & 145 $\pm$ 17 \\
\verb|NFW_sub| & \verb|PJ_free| & 362 &  1.13 $\pm$ 0.03 & 2.81 $\pm$ 0.25 & 150 $\pm$ 18 \\
\verb|PJ_tidal| & \verb|PJ_tidal| & 348 &  1.07 $\pm$ 0.04 & 1.54 $\pm$ 0.07 & 53 $\pm$ 1 \\
\verb|None| & \verb|PJ_free| & 343 &  1.10 $\pm$ 0.04 & 2.41 $\pm$ 0.23 & 123 $\pm$ 18 \\
\verb|PJ_tidal| & \verb|None| & 16 &  - & - & - \\
\verb|PJ_free| & \verb|None| & 16 &  - & - & - \\
\verb|NFW_sub| & \verb|None| & 13 &  - & - & - \\
\verb|None| & \verb|None| & $\equiv 0$ &  - & - & - \\
\botrule
\end{tabular}
\end{table}

\newpage
\section*{Figures}

\setcounter{figure}{0}
\renewcommand{\figurename}{Figure}

\begin{figure}[H]
\centering
\includegraphics{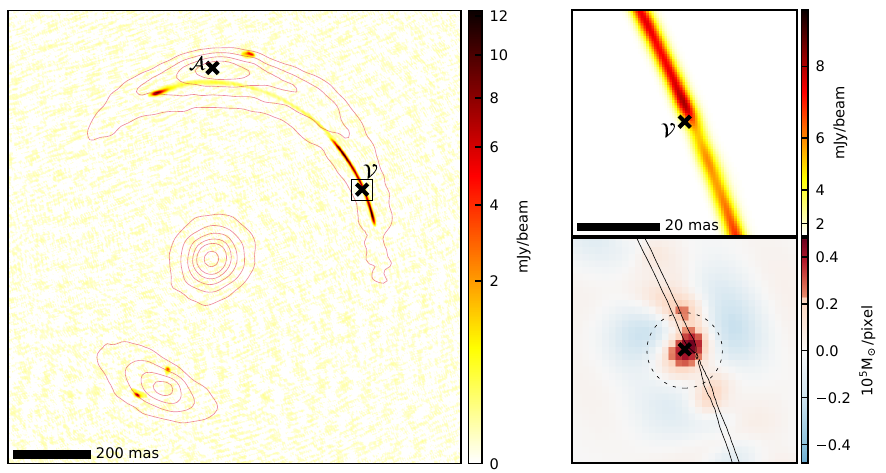}
\caption{The strong gravitational lens system \bnte. \textbf{Left:} Our best model of the 1.7 GHz global VLBI observation used \revisions{here}, which has been re-convolved with the main lobe of the interferometer's PSF and added to the residuals (34 $\mu$Jy/beam RMS). For reference, red contours show a 2.1 $\mu$m observation from the W. M. Keck Observatory adaptive optics system \cite{lagattuta2012}. The positions of two low-mass perturbers are marked with black X's. The $2\times10^8\units{\msun}$ object first detected by \cite{vegetti2012} is labeled $\aodet$, while the $1.13\times 10^6\units{\msun}$ detection reported \revisions{here} is labeled $\vlbidet$. The zoomed-in region shown in the right two panels is indicated by the black square, which has a side length of 60 milli-arcseconds.  \textbf{Top right:} Detail of the bright arc around $\vlbidet$, with the \revisions{colour} scale modified to emphasize the gap in the arc produced by $\vlbidet$'s gravitational perturbation. \textbf{Bottom right:} Gravitational imaging (GI) corrections to the lensing convergence (expressed in units of lens-plane surface mass density), showing a compact, positive feature whose position and mass are consistent with the independent parametric modeling results for $\vlbidet$.  The dashed black circle has a radius of 80 pc, and the \revisions{lensed emission} are indicated by the black contours. } \label{fig:deconvolved}
\end{figure}

\begin{figure}[H]
\centering
\includegraphics{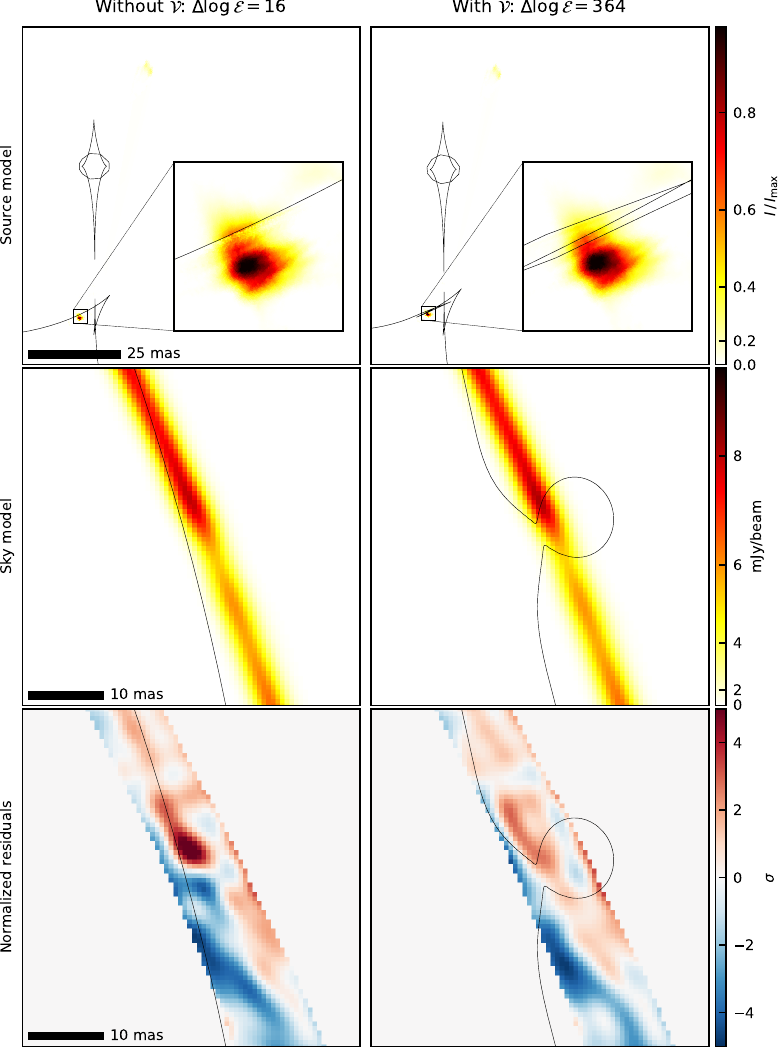}
\caption{Visual comparison of models without and with perturber $\vlbidet$.  \textbf{Top row:} Pixellated source surface brightness maps. In the model without $\vlbidet$ , the source model attempts to fit the gap in the bright lensed arc, resulting in a sharp discontinuity in the surface brightness along the lensing caustic \secondrevisions{(left column)}.  The gravitational effect of $\vlbidet$ folds the caustic over onto the bright radio lobe, correcting this discontinuity and recovering a smooth, contiguous image  \secondrevisions{(right column)}.  The inset region around the bright, \revisions{quadruply-imaged} source component has a side length of 4 milli-arcseconds. \revisions{Also note the presence of a much fainter, doubly-imaged source component $\sim50\units{mas}$ to the north of the bright component.}
\textbf{Middle row:} Lens-plane surface brightness maps (re-convolved with the main lobe of the PSF).  Note that the gap in the arc is still present in the model without $\vlbidet$; in this case the source has attempted to absorb the discontinuity. \textbf{Bottom row:} Residuals, which have been normalized in the visibility plane and Fourier-transformed into the image plane for ease of visualization. The inclusion of $\vlbidet$ corrects a $>5\sigma$ peak in the residuals (corresponding to a $2$ percent change in surface brightness) near the intersection between the critical curve and the lensed arc. The zoomed-in region in the middle and bottom rows is the same as the inset region in Figure \ref{fig:deconvolved}. 
} \label{fig:modelcmp}
\end{figure}

\begin{figure}[H]
\centering
\includegraphics{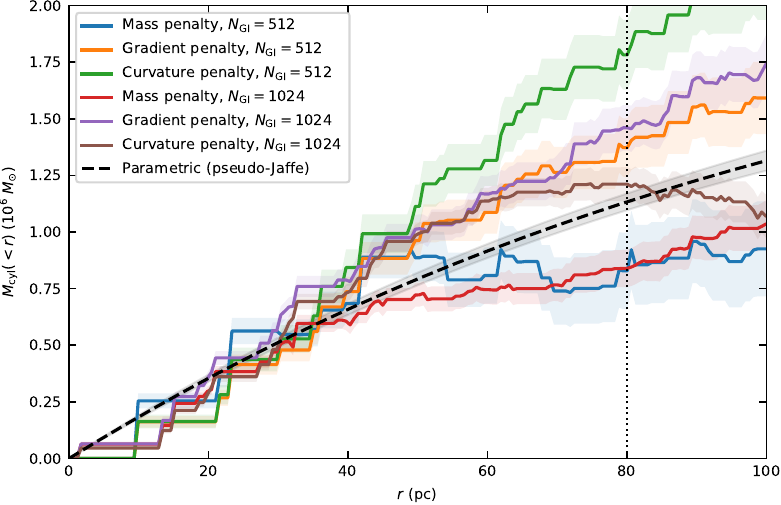}
\caption{Comparison of mass profiles of detection $\vlbidet$, in terms of cylindrical enclosed mass $M_\mathrm{cyl}(<r)$, for both gravitational imaging and parametric modeling procedures. \secondrevisions{For the gravitational imaging models, we compare three different regularization types (penalizing the convergence, gradient of convergence, or curvature of convergence) and two different grid resolutions ($N_\mathrm{GI}=512$ and $N_\mathrm{GI}=1024$, corresponding to pixel sizes of $3.5\units{mas}$ and $1.8\units{mas}$, respectively).} The profiles are consistent to within $\sim 50$ percent at a radius of 80 pc (vertical dashed line). The discrete steps in the GI curves are due to the pixellated nature of the convergence corrections. \secondrevisions{Shaded regions denote the $1\sigma$ uncertainties in the enclosed mass profiles.}. } \label{fig:mencl}
\end{figure}

\newpage
\bibliography{sn-bibliography}% common bib file

\newpage

% \appendix

\section*{Supplementary Figures}

\setcounter{figure}{0}
\renewcommand{\figurename}{Supplementary Figure}

\begin{figure}[h]
\centering
\includegraphics[scale=0.9]{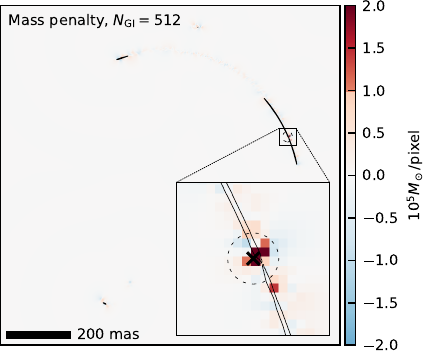}~~~~~~\includegraphics[scale=0.9]{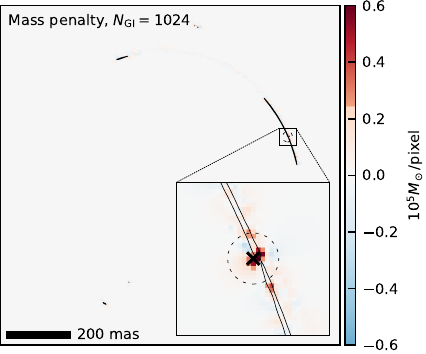}\\
\includegraphics[scale=0.9]{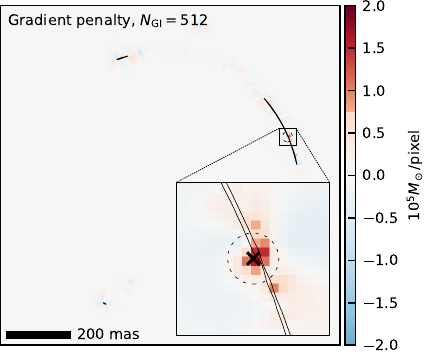}~~~~~~\includegraphics[scale=0.9]{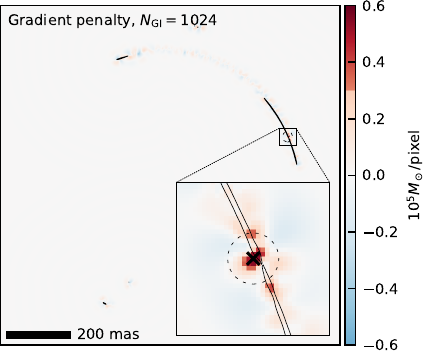}\\
\includegraphics[scale=0.9]{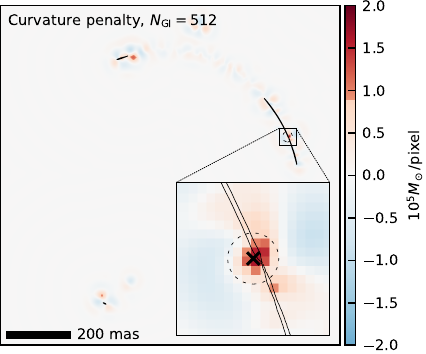}~~~~~~\includegraphics[scale=0.9]{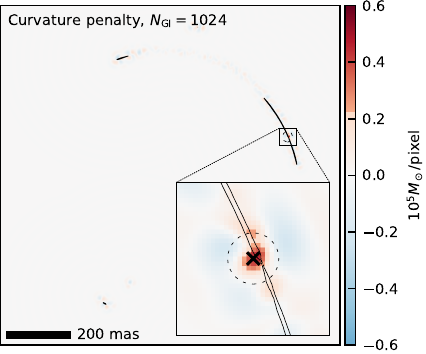}\\
\caption{Non-parametric gravitational imaging (GI) corrections to the lensing convergence (expressed in units of lens-plane surface mass density), for six different combinations of regularization type and grid resolution $N_\mathrm{GI}$. The discontinuities in the intensities of the colour scales indicate the $3\sigma_\mathrm{GI}$ thresholds used to identify features of interest in the convergence corrections. In all six GI runs, we recover compact, positive convergence features in the same location along the bright arc.  These positions are consistent with the independent parametric modeling results for detection $\vlbidet$, whose location is shown by the black X's.  The dashed black circles in the inset panels have a radius of 80 pc. The lensed images are indicated by the black contours.  } \label{fig:px}
\end{figure}

\begin{figure}[h]
\centering
\includegraphics[scale=0.9]{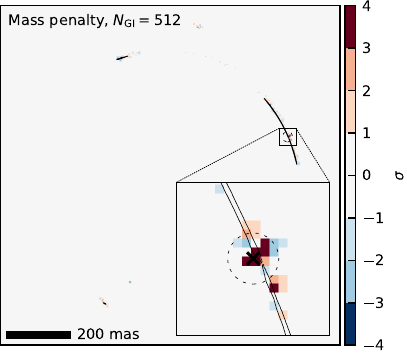}~~~~~~\includegraphics[scale=0.9]{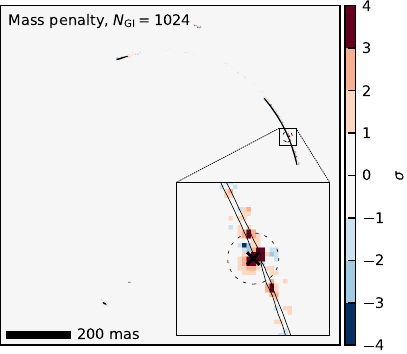}\\
\includegraphics[scale=0.9]{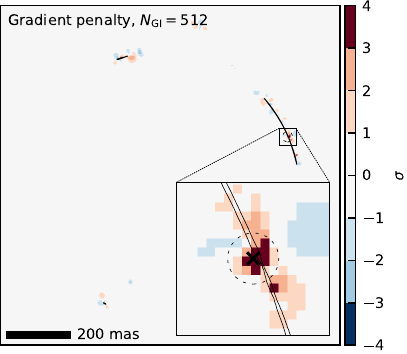}~~~~~~\includegraphics[scale=0.9]{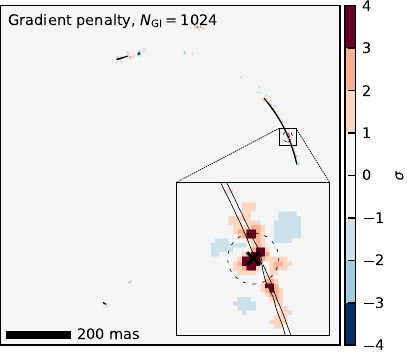}\\
\includegraphics[scale=0.9]{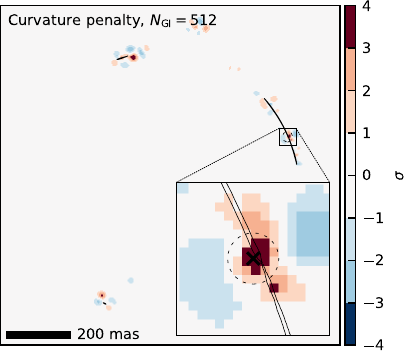}~~~~~~\includegraphics[scale=0.9]{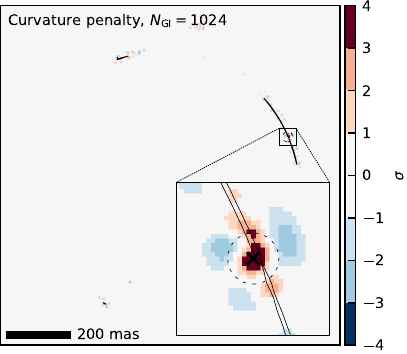}\\
\caption{Significance map for gravitational imaging (GI) corrections to the lensing convergence, expressed in units of $\sigma_\mathrm{GI}$.  While features above the $3\sigma_\mathrm{GI}$ threshold appear at various locations in some of the runs, only the feature associated with detection $\vlbidet$ (black X's) is robust and consistent across all six different combinations of regularization type and grid resolution $N_\mathrm{GI}$. See Supplemental Figure \ref{fig:px} for GI maps in units of surface mass density. } \label{fig:sigma}
\end{figure}

\begin{figure}[h]
\centering
\includegraphics{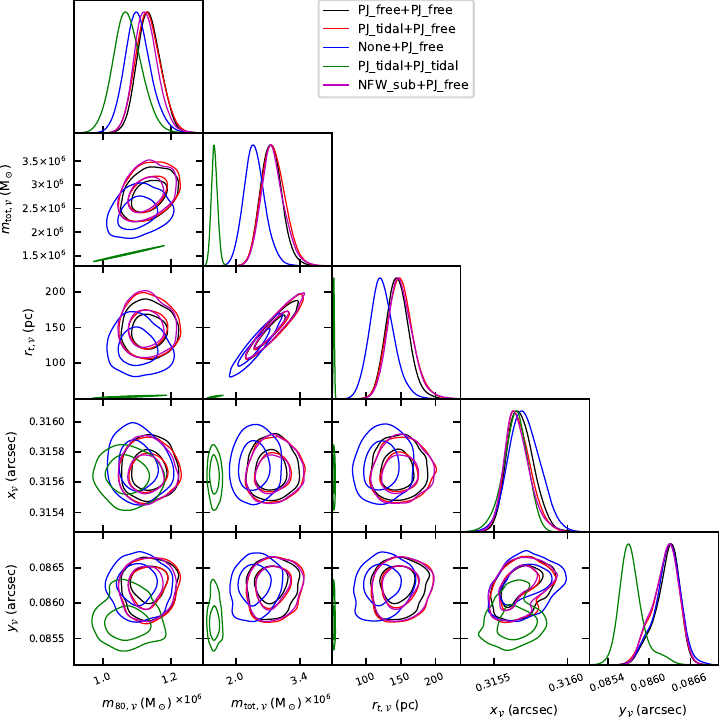}
\cprotect\caption{Parametric modeling results for detection $\vlbidet$. 
$\meightyv$ is the cylindrical mass contained within a radius of 80 pc, where the data are most sensitive to the enclosed mass (see the main text). When the mass of $\vlbidet$ is expressed as $\meightyv$ rather than $m_{\mathrm{tot},\mathcal{V}}$, the properties of $\vlbidet$ are consistent across all models. This indicates that the inferred properties of detection $\vlbidet$ are robust to the model parameterization for detection $\aodet$.  $\meightyv$ is derived from $m_{\mathrm{tot},\mathcal{V}}$ and $r_{t,\mathcal{V}}$, and is not a free parameter. Likewise, the tidal radius $r_{t,\mathcal{V}}$ of $\vlbidet=$ \verb|PJ_tidal| is derived from the lens and perturber masses and positions. }\label{fig:subcorner}
\end{figure}

\begin{figure}[h]
\centering
\includegraphics{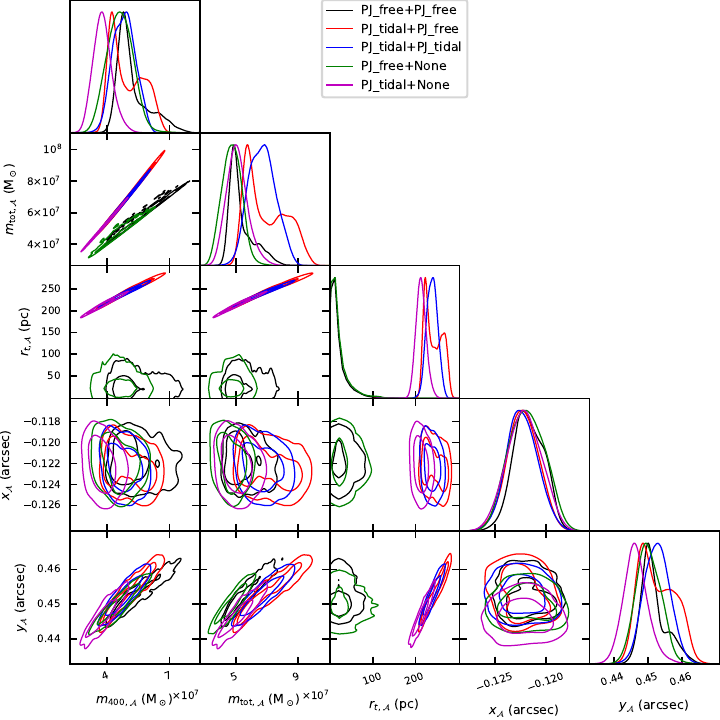}
\cprotect\caption{Parametric modeling results for detection $\aodet$, for runs where $\aodet$ is modeled as pseudo-Jaffe (PJ) subhalo profile. 
$\mfha$ is the cylindrical mass contained within a radius of 400 pc, which is the transverse distance between the position of $\aodet$'s center and the nearest lensed radio emission.  Hence, we expect only $\mfha$ rather than $m_{\mathrm{tot},\mathcal{A}}$ to be robustly constrained by the data, and find that it is consistent between all runs. Meanwhile $r_{t,\aodet}$ is unconstrained and always less than 400 pc, which simply reflects the tidal radius relation or the log-uniform prior. $\mfha$ is roughly consistent with the findings of Despali et al. (see main text), who infer $\mfha=(7.7 \pm 0.1)\times 10^7\units{\msun}$ from a Keck AO observation of \bnte. $\mfha$ is derived from $m_{\mathrm{tot},\mathcal{A}}$ and $r_{t,\mathcal{A}}$, and is not a free parameter. Likewise, the tidal radius $r_{t,\mathcal{A}}$ of $\aodet=$ \verb|PJ_tidal| is derived from the lens and perturber masses and positions.} \label{fig:aopjcorner}
\end{figure}

\begin{figure}[h]
\centering
\includegraphics{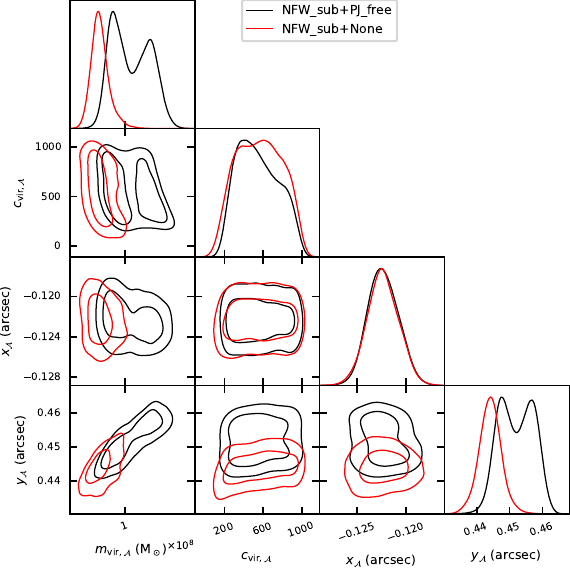}
\caption{Parametric modeling results for detection $\aodet$, for runs where $\aodet$ is modeled as an NFW profile fixed to the lens redshift of $z=0.881$. The virial concentration $c_{\mathrm{vir},\mathcal{A}}$ is unconstrained, reflecting the fact that the  nearest lensed radio emission  lies 400 pc from the center of $\aodet$, and therefore lacks constraining power over the density profile of $\aodet$ within 400 pc. Interestingly, the inclusion of detection $\vlbidet$ in the lens model introduces a second mode in the posterior distribution of $\aodet$'s parameters. This sensitivity to the global lens mass distribution, in contrast with $\vlbidet$, is due to the fact that $\aodet$ does not lie directly on top of lensed radio emission as $\vlbidet$ does. } \label{fig:aonfwcorner}
\end{figure}

\begin{figure}[h]
\centering
\includegraphics{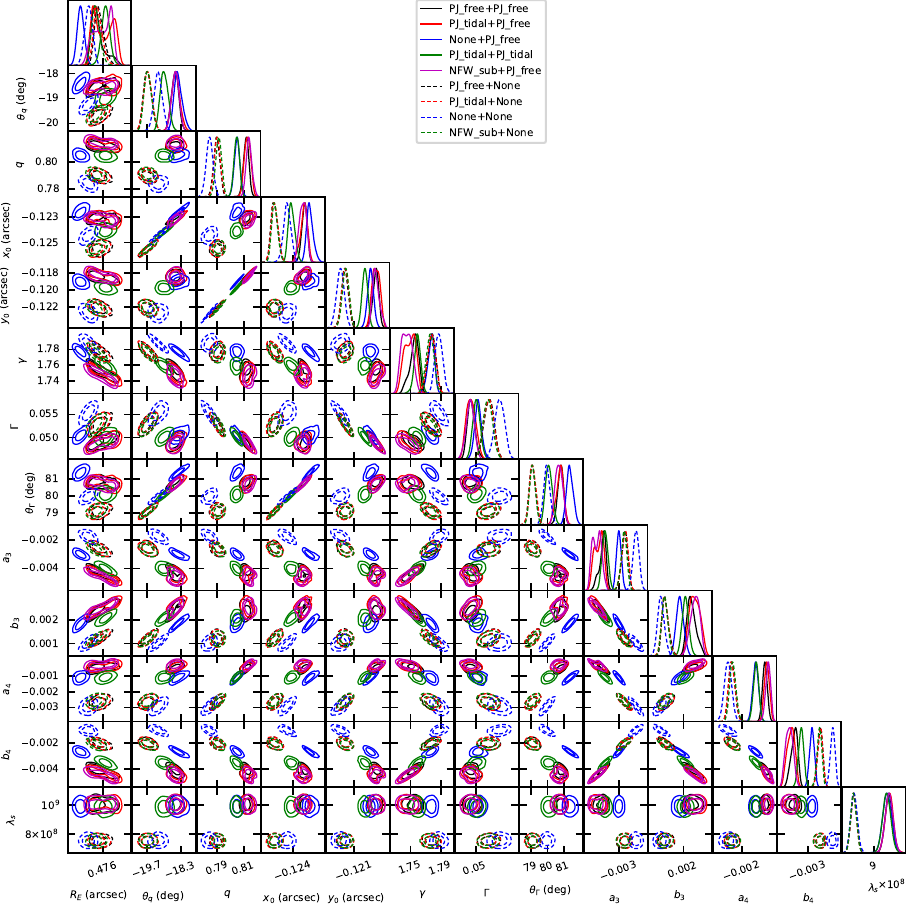}
\caption{Posterior distributions of macro-model parameters for all parametric models. Models including $\vlbidet$ are shown in solid contours, while models without $\vlbidet$ are shown in dashed contours.  Inferred properties of the main lens profile are  mostly sensitive to the presence or absence of $\vlbidet$. In particular, the shape of the main lens (the axis ratio $q$ and multipole coefficients $a_3$, $b_3$, $a_4$, $b_4$) changes depending on whether or not $\vlbidet$ is present.   Most notable, however, is the consistent increase in the preferred value of the source regularization hyper-parameter $\lams$ when $\vlbidet$ is included in the model. This reflects the fact that the source reconstruction is better-focused, and hence smoother, than in the case without $\vlbidet$; see Figure \ref{fig:modelcmp} of the main text. } \label{fig:macrocorner}
\end{figure}

\bigskip

\end{document}